\documentstyle[12pt]{article}

\newcommand{\V}{{\varphi}}
\newcommand{\BE}{\begin{eqnarray}}
\newcommand{\EE}{\end{eqnarray}} 
\newcommand{\NN}{\nonumber}
\begin{document}

{\pagestyle{empty} }

\vskip 6mm

\centerline{\large \bf Generalized Area Law }
\centerline{\large\bf under Multi-parameter Rotating Black Hole Spacetime} 

\vspace{10mm}

\centerline{Masakatsu Kenmoku \footnote{E-mail address:
        kenmoku@asuka.phys.nara-wu.ac.jp}, 
           Yuki Kobayashi \footnote{E-mail address: 
        kobayashi@asuka.phys.nara-wu.ac.jp}}
\centerline{\it Department of Physics,
Nara Women's University, Nara 630-8506, Japan}

\vskip 1cm \centerline{\bf Abstract}

We study the statistical mechanics 
for quantum scalar fields under  
the multi-parameter rotating black hole spacetime 
in arbitrary $D$ dimensions.  
The method of analysis is general in the sense that 
the metric does not depend on the explicit black hole 
solutions. 
The generalized Stefan-Boltzmann's law 
for the scalar field is derived 
by considering the allowed energy region properly. 
Then the generalized area law for the 
scalar field entropy is derived 
by introducing the invariant regularization parameter 
in the Rindler spacetime. 
The derived area law is applied
to Kerr-AdS black holes in four and five dimensions. 
Thermodynamic implication is also discussed. 
 
\vskip 3mm

\noindent PACS number(s): 04.62.+v, 04.70.-s, 04.70.Dy, 
97.60.Lf \hfil \vfill

\newpage

\setcounter{equation}{0}
\renewcommand{\theequation}{\thesection.\arabic{equation}}

\section{Introduction}

Recently many evidences of rotating black holes are observed.  
Such black holes are well described by axisymmetric solutions 
of the Einstein's field equations 
\cite{mtw:,townsend:,stephani:}.  
Furthermore 
rotating black holes in higher dimensions 
\cite{mp:}
and in the (anti)-
de Sitter spacetime of (negative) cosmological term 
have been had great interests theoretically 
in view of the string theory, M-theory and the 
AdS/CFT correspondence \cite{h:,g:,y:}.   

The matter fields may be absorbed into black holes 
by strong gravity and the information of the matter fields 
may transformed to the black holes.  
In order to satisfies the 
first and second laws of thermodynamics, 
black hole themselves have their own entropy  
proportional to the area of the horizon,  
which is known as the black hole thermodynamics 
\cite{smarr:,bekenstein:,bardeen:,gibbons-hawking:}.  
Black holes attract matter fields around them and into their horizon.  
On the other hand black holes may radiate some matter fields 
by the strong gravitational acceleration 
\cite{hawking:,penrose:,cheist:}. 
Therefore it is worthwhile to study the matter filed contribution 
to black hole thermodynamics as a quasi-equilibrium state around black
holes. 

The scalar field contribution to the entropy under  
Schwarzschild spacetime was studied extensively and 
the area law was shown to hold 
\cite{hartle:,thooft:,susskind:,dealwis:,kenmoku:05}. 
Because of the super-radiant stability and/or instability 
\cite{thorne:}, there were some troubles to derive the area law 
for the scalar field under the rotating black hole spacetime 
in (2+1) and (3+1) dimensions.  
For Ba$\tilde{n}$ados-Teitelboim-Zanelli (BTZ) 
black hole spacetime in (2+1) dimensions 
\cite{ichinose:,kimkim:,ho2:,fatibene:}
and Kerr black hole spacetime in (3+1) dimensions 
\cite{lee:,mann:,recent:}, 
some additional divergences appeared in the statistical sum 
of quantum states of the scalar fields, and then the non-rotating limit 
cannot be taken. 
Another difficulty in BTZ spacetime in (2+1) dimension was that 
the result depends on the method of calculation in addition to the above
difficulties. 
We have studied 
the scalar field contribution to the rotating black hole entropy 
in single-parameter case in our previous paper\cite{kenmoku:06}.  
We have taken into account the energy restriction  
for scalar fields in the statistical integration,  
which define the Boltzmann' factor well defined. 
We have also introduced the zenithal angle dependent 
regularization parameter in original spacetime,  
which derive the area law for the scalar field. 
The obtained area law of the entropy is applied 
to BTZ black hole spacetime and Kerr black hole spacetime,  
where two methods are adopted: 
semi-classical method and Euclidean path integral method 
in order to confirm the result. 

In this paper, 
we extend the method in our previous paper \cite{kenmoku:06} to 
multi-parameter rotating black hole spacetime 
in arbitrary $D$ dimensions. 
We try to improve the method to obtain the area law in the following points.
\begin{itemize}
\item[(i)]
Back ground black hole metric is required minimally and 
is in possible general form. 
\item[(ii)]
The energy restriction for scalar fiels will be derived ctj
 rigorously. This is important in performing 
the energy integration in statistical mechanics.  
\item[(iii)] 
The regularization parameter will be defined  
 systematically in the Rindler spacetime. 
By this definition, the regularization parameter will 
become invariant under radial coordinate transformation.    
\end{itemize}

Under these considerations, 
we study the statistical mechanics for 
scalar fields under the multi-parameter rotating 
black hole spacetime 
by adopting the Euclidean path integral method. 
The obtained thermodynamic quantities of scalar fields  
consist of the product of temperature part 
and volume part in the optical space , 
which are in the form of generalized Stefan-Boltzmann's law. 
The main contribution to entropy is from the optical volume 
near the horizon, 
which leads the generalized area law of the entropy naturally.  
Here we introduce the regularization parameter in the Rindler 
spacetime to cut-off the ultra-violet divergence on the horizon. 
The ultra-violet cutoff for non-rotating black holes 
is originally introduced  
in the brick wall model by 't Hoot\cite{thooft:}.  
The generalized area law is applied 
to Kerr-(anti)-de Sitter (Kerr-AdS) black holes in 
four and five dimensions. 
Kerr-AdS spacetime is interesting in view of 
recent development of the AdS/CFT correspondence to understand 
the non-perturbative effects.   
Higher dimensional multi-parameter rotating black holes are also  
interesting to lead the deeper understanding of the black hole
thermodynamics. 
The obtained thermodynamic quantities for scalar fields 
are shown to satisfy 
the thermodynamic (Gibbs-Duhem) relation 
and the first law of thermodynamics. 
It is interesting that they also satisfy 
the thermodynamics in the language of black hole variables. 
This implies that the thermodynamics for scalar fields 
can be consistently considered to be the thermodynamics 
 for the black hole itself. 

The organization of this paper is the following. 
In section 2, the statistical mechanics for the scalar fields 
is studied. 
The generalized Stefan-Boltzmann' law  
is derived in section 2.1 and the generalized area law 
is derived in section 2.2. 
In section 3, the generalized area law is applied to 
Kerr-AdS black hole spacetime in (3+1) and (4+1) dimensions. 
The relation between the thermodynamics for scalar fields  
and that of the black holes is studied in section 4. 
Summary and discussions are given in the last section.

\section{Statistical Mechanics for Scalar Field 
under Multi-parameter Rotating Black Hole Spacetime }

In this section, we study the statistical mechanics 
for the quantum scalar field  
under multi-parameter rotating black hole spacetime 
in arbitrary $D$ dimensions.  
Our method of analysis is general in the sense 
that the metric does not depend on the explicit black hole solutions. 
We first study the generalized Stefan-Boltzmann's law 
by adopting the Euclidean path integral method.  
This method leads to the optical space representation 
for the thermodynamic quantities naturally. 
To derive the generalized area law, 
we evaluate the scalar field entropy 
in the Rindler spacetime near the black hole horizon. 
We adopt units such that $c=\hbar=k_{{\rm B}}=G=1$ unless otherwise specified.

\subsection{Generalized Stefan-Boltzmann's law}

We set the $D$-dimensional polar coordinate as 
\begin{eqnarray}
x^{\mu}=(x^0, x^1, x^2, \cdots , x^{D-1} )
=(t, \V_{1},\cdots, \V_{p} , \theta_{1}, \cdots , \theta_{q}, r)
\ \ , \label{e1}
\end{eqnarray}
where the number of azimuthal angles $\V_{a}$ is $p$ 
and that of zenithal angles $\theta_{m}$ is $q$, where $p+q+2=D$. 
The line element is assumed to be of the form
\begin{eqnarray}
ds^2
&=&\sum_{\mu,\nu=0}^{D-1}g_{\mu\nu}x^{\mu}x^{\nu}\NN\\
&=&g_{tt}dt^2+2\sum_{a=1}^{p}g_{t\V_{a}}dtd\V_{a} 
+\sum_{i,j=1}^{D-1} h_{ij}dx^{i}x^{j}\NN\\
&=&g_{tt}dt^2+2\sum_{a=1}^{p}g_{t\V_{a}}dtd\V_{a} 
+\sum_{a,b=1}^{p}g_{\V_{a}\V_{b}}d\V_{a}d\V_{b} \NN\\
&&
+\sum_{m,n=1}^{q}g_{\theta_{m}\theta_{n}}d\theta_{m}d\theta_{n}+g_{rr}dr^2
\ ,  \label{e2}
\end{eqnarray}
where special metric components $h_{ij}$ in the second line 
are defined as
\begin{eqnarray}
h_{ij}:=(g_{\V_{a}\V_{b}}, g_{\theta_{m}\theta_{n}}, g_{rr}) \ ,
\label{e3}
\end{eqnarray}
where they are already in the black diagonal form 
with respect to $\V_{a}$, $\theta_{m}$ and $r$. 
Our assumptions on the metric are the followings:  
\begin{itemize}
\item[$\Diamond$]Metric Condition 1 :\\
The off-diagonal metrices related to the time component 
are those between time and azimuthal components 
$g_{t\V_{a}}$  in the second terms of Eq.(\ref{e2}). 

\item[$\Diamond$]Metric Condition 2:\\
All metrices are assumed to be functions 
of radial variable $r$ and zenithal angles $\theta$ 
and do not depend on time $t$ and azimuthal angles $\V_{a}$.    
\end{itemize}
We use the suffix notation: $\mu,\nu=0, \cdots , D-1 $ for 
full $D$ dimensional spacetime, 
$i,j=1, \cdots , D-1$ for special components 
$a,b=1, \cdots , p$ for azimuthal angle components $(\V)$
and $m,n=1, \cdots , q$ for zenithal angle components $(\theta)$ 
in the followings.
 

With the metric in Eq.(\ref{e2}),  
the matter action for the scalar field $\Phi$ of mass $m$ in $D$
 dimension is 
\begin{eqnarray}
&&I_{\rm scalar}=\int d^Dx\sqrt{-g} \ {\mathcal{L}}_{\rm scalar}(x)\nonumber\\
&&{\mathcal{L}}_{\rm scalar}(x)= -\frac{1}{2} 
(g^{\mu\nu}\partial_{\mu}\Phi(x)\partial_{\nu}\Phi(x)
+m^2\Phi(x)^2) \label{e4}
\end{eqnarray}
The canonical momentum of the scalar field is defined by 
\begin{eqnarray}
\Pi(x):=\frac{\partial {\mathcal{L}}_{\rm scalar}(x)}
{\partial \partial_{t}{\Phi}(x)}
=-g^{t t}\partial_{t}{\Phi}(x)
-\sum_{a=1}^{p}g^{t\V_{a}}\partial_{\V_{a}}\Phi(x) 
\ , \label{e5}
\end{eqnarray}
and the quantization condition is given by 
\begin{eqnarray}
\left[\Phi(x), \Pi(y) \right]\mid_{t=t'}
=i\frac{\delta^{D-1}({x}-{y})}{\sqrt{-g}}
\ . \label{e6}
\end{eqnarray}


Because the metric components in Eq.(\ref{e2}) are assumed not to
depend on $t$ and $\V_{a} \ (a=1, \cdots, p)$,  
there exist $(p+1)$ Killing vectors:    
\begin{eqnarray}
\xi_{t}=\partial_{t}\ ,  \ 
\xi_{\V_{a}}=\partial_{\V_{a}}\ \ \ (a=1, \cdots, p)\ .  
\label{e7}
\end{eqnarray}
The existence of the Killing vectors implies the conservations of 
the total energy $H$ 
and the azimuthal angular momentum $P_{\V_{a}}$ of 
the scalar field, which are defined by  
\begin{eqnarray}
H:&=-&\int_{\Sigma} (\xi_{t})_{\mu}{\mathcal{T}}^{\mu t} d \Sigma_{t}
=\int d^{D-1}x \sqrt{-g}
\left(\Pi \partial_{t}{\Phi}-\mathcal{L}_{\rm scalar}\right)
\ , \nonumber \\ 
P_{\V_{a}}:&=&\int_{\Sigma}(\xi_{\V_{a}})_{\mu}
{\mathcal T}^{\mu t} d \Sigma_{t} 
= \int d^{D-1}x \sqrt{-g}\, \left(-\Pi\, \partial_{\V_{a}}{\Phi}\right)
\ ,  \label{e8}
\end{eqnarray}
where the energy-momentum tensor is    
${\mathcal T}^{\mu\nu}=-({2}/{\sqrt{-g}}) 
({\delta I_{\rm scalar}}/{\delta g_{\mu\nu} })$.
A new Killing vector $\eta$ is introduced combining 
(p+1) Killing vectors in Eq.(\ref{e7}) linearly as 
\BE
\eta:=\xi_{t}+\sum_{a=1}^{p}\Omega_{{\rm H}{a}}\xi_{\V_{a}}\ . 
\label{e9} 
\EE
Its quadratic form satisfies the identical relation: 
\BE
\eta^2
&=& g_{tt}+2\sum_{a=1}^{p}g_{t\V_{a}}\Omega_{{\rm H}{a}}
+\sum_{a,b=1}^{p}g_{\V_{a}\V_{b}}\Omega_{{\rm H}{a}}\Omega_{{\rm H}{b}}
\NN\\
&=&\frac{1}{g^{tt}}
+\sum_{a,b=1}^{p}g_{\V_{a}\V_{b}}(\Omega_{{\rm H}{a}}-\Omega_{a})
(\Omega_{{\rm H}{b}}-\Omega_{b})\ ,
\label{e10}
\EE
where $\Omega_{a}$ and $\Omega_{{\rm H}{a}}$ are the angular velocities 
with respect to azimuthal angles $\V_{a}$ at an arbitrary 
position $x^{\mu}$ and at the black hole horizon $r=r_{\rm H}$ as
\BE
\Omega_{a}:=\frac{g^{t\V_{a}}}{g^{tt}} \ \  \ , \  \ \ 
\Omega_{{\rm H}a}:=\left. \frac{g^{t\V_{a}}}{g^{tt}}
\right|_{r=r_{\rm H}} \ \ \ (a=1, \cdots, p) \ .  
\label{e11}
\EE 
The derivation of the identity equation (\ref{e10}) is given in appendix A.  
This equation shows that the new Killing vector $\eta$ is null 
and is future directed at the horizon.    
Corresponding to the new Killing vector $\eta$, 
a new conserved quantity is introduced combining 
the energy and the angular momenta 
in Eq.(\ref{e8}) as    
\begin{eqnarray}
H - \sum_{a=1}^{p}\Omega_{\V_{a}}P_{\V_{a}}   
= - \int_{\Sigma} \eta^{\mu}T_{\mu t} d \Sigma^{t} 
=:  \int d^{D-1}x \sqrt{-g} \ \mathcal{H}'  \ ,
\label{e12}
\end{eqnarray}
where the newly defined Hamiltonian density $\mathcal{H}'$ is given by 
\begin{eqnarray}
{\mathcal{H}}' 
&=&
\Pi \partial_{t}{\Phi}-{\mathcal{L}}_{\rm scalar}
+\sum_{a=1}^{p}\Omega_{\V_{a}}\Pi\partial_{\V_{a}\Phi} \NN\\
&=&\frac{1}{2}(
-\frac{\Pi^2}{g^{tt}}
 + \sum_{i,j=1}^{D-1}{{h}^{ij}}
\partial_{i} \Phi \partial_{j} \Phi 
+m^2\Phi ^2)
+\sum_{a=1}^{p}(\Omega_{{\rm H}a}-\Omega_{a})\Pi\partial_{\V_{a}}\Phi
,  \label{e13}
\end{eqnarray}
where $h^{ij}$ denotes   
the contravariant component of the special metric $h_{ij}$ 
in Eq.(\ref{e3}):  
\BE
 \sum_{j=1}^{D-1}h_{ij}{h}^{jk}=\delta_{i}^{k}
\ \  {\mbox with}\ \  
h^{ij}:=(g^{\V_{a}\V_{b}}-\frac{g^{t\V_{a}}g^{t\V_{b}}}{g^{tt}}, 
g^{\theta_{m}\theta_{n}}, g^{rr} )
\ .
\label{e14}
\EE 
The derivation of the expression of Eq.(\ref{e13}) is given in appendix B. 
The expression (\ref{e13}) shows that 
the newly introduced energy 
${H}-\sum_{a=1}^{p}\Omega_{H_{a}}P_{\varphi_{a}}$ is positive definite 
near the horizon $r\simeq r_{\rm H}$:  
\begin{eqnarray}
H-\sum_{a=1}^{p}\Omega_{H_{a}}P_{\phi_{a}}
=\int d^{D-1}x\sqrt{-g}\,{\mathcal{H'}}  
\geq 0 \ . 
\label{e14.5}   
\end{eqnarray}
In the following, we adopt the near horizon approximation 
$\Omega_{{\rm H}a}\simeq\Omega_{a}$ near $r\simeq r_{H}$, that is,  
the cross term in $\mathcal{H}'$ in Eq.(\ref{e13}) is neglected and 
the energy restriction (\ref{e14.5}) holds.  


Next we consider the partition function of the temperature $T=1/\beta$ 
as 
\begin{eqnarray}
Z={\rm Tr} [\exp{(-\beta(H-\sum_{a=1}^{p}\Omega_{{\rm
 H}a}P_{\V_{a}}))}]\ .    
\label{e15}
\end{eqnarray}
The exponent of the Boltzmann factor 
$H-\sum_{a=1}^{p}\Omega_{{\rm H}a}P_{\V_{a}}$ is understood taking
account of the multi-parameter rotation effect according to
the Hartle-Hawking vacuum \cite{hartle:} and 
is positive definite near the horizon 
as shown in Eq.(\ref{e13}), which ensure 
the partition function to be well defined. 
In order to calculate the partition function, 
we express it in the Euclidean path integral form as
\begin{eqnarray}
Z=\int[{\mathcal D} \Phi {\mathcal D}\Pi g_{E}^{1/2}]\exp \left(
\int_{0}^{\beta}d\tau \int d^{D-1} x \sqrt{g_{E}}
(i\Pi\partial_{\tau}{\Phi}-\mathcal{H}')
\right)\ ,    
\label{e16} 
\end{eqnarray}
where the Euclidean time $\tau=it$, the Euclidean metric 
$g_{\tau\tau}=-g_{tt}$, $g_{\tau\V_{a}}=-ig_{t\V_{a}}$ 
and its determinant $g_{E}$ are used. 
The periodic boundary condition for the scalar field 
is required in the integration: $\Phi(x,\tau=\beta)=\Phi(x,\tau=0)$. 

After performing the momentum field $\Pi$ integration and the integration by
parts,  the partition function becomes  
\begin{eqnarray}
Z=\int [{\mathcal{D}} \Phi g_{E}^{1/4} (g^{\tau\tau})^{1/2}]
\exp{(
-\int_{0}^{\beta}d\tau \int d^{D-1}x \ , \frac{\sqrt{g_{E}} g^{\tau\tau} }{2}
\Phi\, \bar{K}\, \Phi}) \ , \label{e17}
\end{eqnarray}
where $\bar{K}$ denotes the kernel 
\begin{eqnarray}
\bar{K}
&:=& - \partial_{\tau}^2 -
\frac{1}{g^{\tau\tau}}(
\sum_{i,j=1}^{D-1}\frac{1}{\sqrt{h}}\partial_{i}(\sqrt{h}h^{ij}\partial_{j}) 
-m^2) 
 \ ,  \label{e18}
\end{eqnarray}
in the optical space of the metric: 
\begin{eqnarray}
d\bar{s}^2=d\tau^2+g^{\tau\tau}\sum_{i,j=1}^{D-1}h_{ij}dx^i dx^j\ .   
\label{e19}
\end{eqnarray}
In Eq.(\ref{e18}), the determinant of $h_{ij}$ is denoted by $h$. 
We have ignored the cross term for the heat kernel $\bar{K}$ in Eq.(\ref{e18})
corresponding to 
that for the new Hamiltonian density $\mathcal{H}'$ in Eq.(\ref{e13}) 
under the near horizon approximation. 

After performing the Gaussian integration with respect to $\Phi$, 
the free energy is obtained using the heat kernel representation 
\cite{birrell:} as
\begin{eqnarray}
\beta F =-\ln Z
=- \frac{1}{2} \mbox{Tr} 
\int_{0}^{\infty}\frac{ds}{s}\exp{(-s\bar{K})}
 \ . \label{e20}
\end{eqnarray}
The trace of the heat kernel is consist of two parts: 
the Euclidean time part and the space part.  
The Euclidean time part is evaluated by using the one dimensional  
eigenfunction of $-i\partial_{\tau}$ as 
\begin{eqnarray}
{\rm Tr} \ \exp{(s\, \partial _{\tau}^2)} &=& 
\int_{0}^{\beta} d\tau \sum_{\ell=-\infty}^{\infty}\frac{1}{\beta}
\exp{(-s(\frac{2\pi \ell}{\beta})^2)} \nonumber \\
&=& \sum_{n=-\infty}^{\infty}\frac{\beta}{(4\pi s)^{1/2}}
\exp{(-\frac{\beta^2 n^2}{4s})} \  . \label{e21}
\end{eqnarray}
The Poisson's summation formula is used in the second equality.  
The space part of the trace is calculated by the asymptotic expansion 
method 
\begin{eqnarray}
&&\mbox{Tr}\, \exp{\left( \frac{s}{g^{\tau\tau}} 
(\sum_{i,j=1}^{D-1}\frac{1}{\sqrt{h}}\partial_{i}(\sqrt{h}h^{ij}\partial_{j})
-m^2)\right)}
 \nonumber \\
&=&\frac{1}{(4\pi s)^{(D-1)/2}}
\sum_{k=0}^{\infty}\bar{B}_{k}{(-s)^k}
\exp{(-\frac{s m^2}{g^{\tau\tau}})} 
\ , \label{e22}
\end{eqnarray}
where $\bar{B}_{k}$'s are the coefficient functions   
of the asymptotic expansion. 
The explicit lower contributions are 
\begin{eqnarray}
\bar{B}_{0}&=& 
\int d^{D-1}x (g^{\tau\tau})^{(D-1)/2}h^{1/2}\ ,
\NN\\ \bar{B}_{1}&=& 
(\frac{1}{4}\frac{D-2}{D-1}-\frac{1}{6})
\int d^{D-1}x (g^{\tau\tau})^{(D-1)/2}h^{1/2}{\bar R}
\ , \label{e23}
\end{eqnarray}
where bar notation denotes the quantities in the optical space. 
\footnote{It is worthwhile to note that 
the lowest order contribution $\bar{B}_{0} $
in Eq.(\ref{e23}) can also be derived 
by using the semi-classical momentum eigenfunction.}

The free energy is expressed by multiplying the two trace parts 
in the lowest order as 
\begin{eqnarray}
F=&-&\int_{0}^{\infty}\frac{ds}{s}\frac{1}{(4\pi s)^{D/2}}
\sum_{n=1}^{\infty}
\exp{(-\frac{\beta^2n^2}{4s})} \bar{B}_{0} 
\exp{(-\frac{m^2 s}{g^{\tau\tau}})} \nonumber\\
=
&-&\frac{1}{\beta^D \pi^{D/2}}
\int_{0}^{\infty}\frac{dt}{t}t^{D/2}{\rm e}^{-t}
\sum_{n=1}^{\infty}\frac{1}{n^D} \bar{B}_{0}
\exp{(-\frac{m^2\beta^2n^2}{4tg^{\tau\tau}})} \ , 
\label{e24}
\end{eqnarray}
where the integration variable  
has been changed in second equality as  
$t={\beta^2n^2}/(4s)$. 
The vacuum energy term $(n=0)$ 
is subtracted in the sum in Eq.(\ref{e24}). 
The contribution of the scalar field mass term 
shown to be very small near the horizon. 
After neglecting the mass term,     
a compact expression for the free energy is obtained   
\begin{eqnarray}
F=-\frac{\zeta(D)\Gamma(D/2)}{\pi^{D/2}\beta^{D}}\ \bar{V}
\ , \label{e25}
\end{eqnarray}
where $\bar{V}$ denotes the volume of the optical space 
defined by  
\begin{eqnarray}
\bar{V} =\bar{B}_{0}\times 1 
=\int d^{D-1}x (g^{\tau\tau})^{(D-1)/2}h^{1/2} \ .
\label{e26}
\end{eqnarray}
The entropy and the internal energy are obtained by 
\begin{eqnarray}
S&=&-\beta^2\frac{\partial F}{\partial \beta}=
\frac{\zeta(D)D\Gamma(D/2)}{\pi^{D/2}\beta^{D-1}}\ \bar{V}
\label{e27}\ , \\
U&=&F+\beta^{-1}S 
=\frac{\zeta(D)(D-1)\Gamma(D/2)}{\pi^{D/2}\beta^{D}}\ \bar{V} 
\label{e28}\ .
\end{eqnarray}
 
The obtained thermal quantities: free energy, 
entropy and internal energy 
in  multi-parameter rotating case are  
apparently same form with that obtained 
in non-rotating \cite{dealwis:} or 
single-parameter rotating cases \cite{kenmoku:06}. 
The multi-parameter effects are included in the inverse temperature 
$\beta$ and the optical volume $\bar V$, which will be evaluated in the 
next subsection. 

The obtained thermal quantities 
in four-dimensional flat spacetime 
are coincide with the Stefan-Boltzmann's law and therefore 
are considered as the generalization of them 
to $D$-dimensional multi-parameter rotating spacetime. 

\subsection{Generalized area law }

In this subsection, 
we evaluate the generalized entropy formula (\ref{e27}) 
to the case of the multi-parameter rotation black holes.  
We first define the horizon and introduce Rindler space 
to evaluate the temperature and to define the 
invariant regularization parameter.  
The dominant contribution to the entropy comes from 
the near horizon region of the optical volume, 
which lead to the area law in general form. 
 

\begin{itemize}

\item[$\Diamond $]Metric Condition 3
:\\
The horizon $r_{\rm H}$ is defined as the position of simple zeros 
for the inverse Euclidean metric components 
$1/g^{\tau\tau}$ and $1/g_{rr}$:   
\begin{eqnarray}
\frac{1}{g^{\tau\tau}}\simeq 
 \left. \partial_{r}\frac{1}{g^{\tau\tau}(\theta)}\right|_{r_{H}}
\times (r-r_{H})
\ \ , \ \  
\frac{1}{g_{rr}}\simeq 
\left. \partial_{r}\frac{1}{g_{rr}(\theta)}\right|_{r_{H}}
\times (r-r_{H})
 \ .   \label{e29}
\end{eqnarray}
For outer horizon 
these coefficients are required to be positive, 
which exclude the cosmological horizon. 
\end{itemize}

Next we set up to introduce Rindler space. 
We rewrite the Euclidean line element in order not to appear 
the cross term between time and azimuthal angle parts explicitly as 
\begin{eqnarray}
ds^2
&=&
g_{\tau\tau}{d\tau}^2+2\sum_{a=1}^{p}g_{\tau\V_{a}}d\tau d\V_{a} 
+\sum_{m,n=1}^{q}g_{\theta_{m}\theta_{n}}d\theta_{m}d\theta_{n}
+g_{rr}dr^2\ , \NN\\
&=&\frac{1}{g^{\tau\tau}}d\tau^2
+\sum_{a,b=1}^{p}
g_{\V_{a}\V_{b}}(d\V_{a}-\Omega_{a}d\tau)(d\V_{b}-\Omega_{b}d\tau)\NN\\
&&+\sum_{m,n=1}^{q}g_{\theta_{m}\theta_{n}}d\theta_{m}d\theta_{n}
+g_{rr}dr^2
\ .  \label{e31}
\end{eqnarray}
The derivation of Eq.(\ref{e31}) is given in appendix C. 
The line element is now in the block diagonal form 
with respect to new azimuthal angles $d\V_{a}-\Omega_{a}d\tau$. 
We introduce a new radial variable $R$, 
which is invariant under the radial coordinate transformation, 
defined by 
\begin{eqnarray}
R(r):=\int_{r_{\rm H}}^{r}dr \sqrt{g_{rr}}\ . 
\label{e32}
\end{eqnarray}
Using the new radial coordinate $R$,  
the line element (\ref{e31}) 
is expressed near the horizon in the form of 
Rindler spacetime as  
\begin{eqnarray}
ds^2\simeq 
\frac{1}{4} 
 \left. \partial_{r}\frac{1}{g^{\tau\tau}(\theta)} 
 \partial_{r}\frac{1}{g_{rr}(\theta)}\right|_{r_{H}}  
{R^2} d\tau^2
+dR^2+(\V, \theta \ {\rm terms}) 
\label{e34} \ .
\end{eqnarray}

 
\begin{itemize}
\item[$\Diamond$]Metric Condition 4 
:\\
The temperature on the horizon $(T_{H}=1/\beta_{H})$ is defined  
by the conditions 
that no conical singularity and no angle dependence 
are required in Rindler space: 
\begin{eqnarray}
\frac{2\pi}{\beta_{H}}
=\left(
\left.
\frac{1}{4}\  
\partial_{r}\frac{1}{g^{\tau\tau}(\theta)}\ 
\partial_{r}\frac{1}{g_{rr}(\theta)}\right|_{r_{H}}
\right)^{1/2} 
=\mbox{independent on}\ \theta
\ . \label{e35}
\end{eqnarray}
    
\end{itemize}


Under the definitions of the horizon (\ref{e29}) and 
the temperature (\ref{e35}), 
we consider the volume of the optical space $\bar V$ in Eq.(\ref{e26}).  
As $\bar V$ is divergent on the horizon, 
we introduce the invariant regularization parameter 
for the small distance $\epsilon_{\rm inv}$ 
and the large distance $L_{\rm inv}$ 
in the Rindler spacetime as  
\begin{eqnarray}
\epsilon_{\rm inv}
:=\int_{r_{H}}^{r_{H}+\epsilon(\theta)}dr\, 
(g_{rr}(\theta))^{1/2}
\ \ , \ \ 
L_{\rm inv}
:=\int_{r_{H}}^{r_{\rm H}+L(\theta)}dr\, (g_{rr}(\theta))^{1/2}
\ . \label{e36}
\end{eqnarray}
The regularization parameters 
in the Rindler spacetime $\epsilon_{\rm inv}, L_{\rm inv}$ 
are required angle independent,  
while those in the original spacetime 
$\epsilon(\theta), L(\theta)$ become angle dependent.   
This is 
because the physical quantities in Rindler space  
should be angle independent just as the temperature,  
for they are invariant under the radial coordinate transformation 
and then observer independent. 
\footnote{Note that the regularization parameter is angle independent 
for non-rotating case in the brick wall model \cite{thooft:}.}
Using these regularization parameters, 
the radial integration for the optical volume $\bar V$ 
is evaluated via Rindler space near the horizon as
\begin{eqnarray}
&&\left. \int_{r_{H}+\epsilon(\theta)}^{r_{\rm H}+L(\theta)}dr\,
 (g^{\tau\tau})^{D/2-1/2}(g_{rr})^{1/2}\right|_{r_{\rm H}}
\NN\\
=&&\left. 
\int_{\epsilon_{\rm inv}}^{L_{\rm inv}}dR\, 
(g^{\tau\tau})^{D/2-1/2}\right|_{r_{\rm H}}
\simeq 
(\frac{\beta_{H}}{2\pi})^{D/2-1/2}\frac{1}{(D-2)\epsilon_{\rm inv}^{\ \ D-2}}
\ ,   \label{e37}
\end{eqnarray}
where the temperature condition (\ref{e35}) is used. 
The contribution from the large regularization parameter $(L_{\rm inv})$ 
is negligible under the magnitude condition:  
$\epsilon_{\rm inv} \ll L_{\rm inv}$. 
The angular integration in $\bar V$ gives the area on the horizon as 
\begin{eqnarray}
A_{H}:=\left. \int \prod_{a=1}^{p} d\V_{a} \prod_{m=1}^{q}d\theta_{m}
\sqrt{{\rm det}(g_{\V_{a}\V_{b}}) \ {\rm det}(g_{\theta_{m}\theta_{n}}})
 \right|_{r_{H}}\ .   \label{e38}
\end{eqnarray}
Combining together the radial coordinate integration (\ref{e37}) 
and the angular integration (\ref{e38}), 
the optical volume $\bar V$ (\ref{e26}) is evaluated as
\begin{eqnarray}
\bar{V}=(\frac{\beta_{\rm H}}{2\pi})^{D-1}
\frac{A_{H}}{(D-2)\epsilon_{\rm inv}^{\ \ D-2}}
\ .   \label{e39}
\end{eqnarray}
Then the generalized entropy formula (\ref{e27}) is evaluated 
near the horizon using Eq.(\ref{e39}) as
\begin{eqnarray}
S= \frac{\zeta(D)D\Gamma(D/2-1)}{2^{D}\pi^{3D/2-1}}
\frac{A_{H}}{{\epsilon_{\rm inv}}^{ D-2}}
\ .  \label{e40}
\end{eqnarray}

We have obtained the generalized area law for 
quantum scalar field under the multi-parameter 
rotating black hole spacetime without using the 
explicit form of black hole solutions. 
The expression of the area law under multi-parameter cases 
has the same form as that under non-rotating  
and single-parameter rotating cases \cite{kenmoku:06}. 
Therefore we can take smooth non-rotating limit of the 
generalized are law. 
The rotating effects are included in the temperature 
and the volume of the optical space. 

The thermal quantities; entropy $S$, internal energy $U$ and 
Helmholtz's free energy $F$ are 
consist of the product form of the temperature ($\beta$) term 
and the volume term $(\bar V)$ in the optical space. 
The contribution of the real space integration in $\bar{V}$ 
is extremely large near the horizon region and then  
the area law of the black hole entropy is derived. 

We also note that the super-radiant modes are take into account 
in our entropy expression (\ref{e40}), because 
the scalar field energy $E$ can be negative if the 
angular momentum $P_{\V_{a}}$ be negative    
and satisfies our energy restriction (\ref{e14.5}): 
$H-\sum_{a=1}^{p}\Omega_{a}P_{\V_{a}}\geq 0$.  

\setcounter{equation}{0}
\section{Application}

In this section,  we apply the generalized 
area law under the multi-parameter rotating black hole 
spacetime  
to the cases of Kerr-AdS black hole spacetime, 
which are interesting and developing recently.  
We first study the four-dimensional Kerr-AdS spacetime  
as single-parameter rotating case and next 
the five-dimensional Kerr-AdS spacetime as 
double-parameter rotating case.  
A special feature in the Kerr-(anti)-de Sitter spacetime is that 
the rotation effects remain even in the asymptotic region 
for rotating black holes.  
In general $D$ dimension, 
the multi-parameter rotating black hole solutions 
are known to exist systematically for the even- and odd-dimensional cases 
respectively.  
The examples in this section may give the understanding for 
the general multi-parameter rotating black hole cases. 
 
\subsection{Kerr-AdS black holes in four dimensions}

The four-dimensional Kerr-AdS metric is given by Carter  
\cite{carter:} as 
\begin{eqnarray}
ds_{D=4}^2=
&-&\frac{\Delta}{\rho^2}
( dt-\frac{a\sin^2{\theta}}{\Xi}d\V )^2 
+\frac{\rho^2}{\Delta}dr^2
+\frac{\rho^2}{\Delta_{\theta}}d\theta^2\NN\\
&+&\frac{\Delta_{\theta}\sin^2{\theta}}{\rho^2}
(a dt-\frac{r^2+a^2}{\Xi}d\V )^2 \ , 
\label{f1}
\end{eqnarray}
with  
\begin{eqnarray}
\Delta&=&(r^2+a^2)(1+r^2\ell^2)-2Mr \ , \ \ 
\Delta_{\theta}=1-a^2\ell^2\cos^2{\theta} \ , \NN\\
\rho^2&=&r^2+a^2\cos^2{\theta}\ \ , \ \ \Xi=1-a^2\ell^2 \ ,  
\label{f2}
\end{eqnarray}
where $M$ , $a$ and $\Lambda=-3\ell^{-2}$ 
denote the black hole mass, its angular momentum per unit mass 
and the cosmological term respectively. 
In order to apply the general area law (\ref{e40}), 
the metric (\ref{f1}) is rewritten in the general form:   
\begin{eqnarray}
ds_{D=4}^2
&=&g_{tt}dt^2+2g_{t\V}dtd\V +g_{\V\V}d\V^2 
+g_{\theta\theta}d\theta^2+g_{rr}dr^2
\ ,  \label{f3}
\end{eqnarray}
where each components are given by 
\begin{eqnarray}
g_{tt}&=&\frac{1}{\rho^2}( -\Delta
+\Delta_{\theta}a^2\sin^2{\theta})\ , \ \
g_{t\V}=\frac{a\sin^2{\theta}}{\rho^2\Xi}
( \Delta -(r^2+a^2)\Delta_{\theta})\ , \NN\\ 
g_{\V\V}&=&\frac{\sin^2{\theta}}{\rho^2\Xi^2}
( -\Delta a^2\sin^2{\theta}+(r^2+a^2)^2\Delta_{\theta} )
\ , \  g_{\theta\theta}=\frac{\rho^2}{\Delta}\ , \   
g_{rr}=\frac{\rho^2}{\Delta_{\theta}} \ .  \label{f4} 
\end{eqnarray} 
In order to obtain the temperature and the angular velocity 
we need to know some of the inverse components of the metric: 
\begin{eqnarray} 
g^{tt}=\frac{g_{\V\V}}{\Gamma}\ , \ \ g^{\V\V}=\frac{g_{tt}}{\Gamma}
\ , \ \ g^{t\V}=-\frac{g_{t\V}}{\Gamma} \ , \label{f5}
\end{eqnarray}
with 
$\Gamma:=g_{tt}g_{\V\V}-g_{t\V}^2
=-\Delta\Delta_{\theta}\sin^2{\theta}/\Xi^2$.
The horizon $r_{H}$ is defined as the larger zero of $1/g^{tt}$ as well as 
$1/g_{rr}$, which gives $\Delta(r=r_{H})=0$. 
The angular velocity and the temperature on the horizon are calculated as  
\begin{eqnarray}
\Omega_{H}&=&\left. \frac{g^{t\V}}{g^{\V\V}}\right|_{r_{H}}
=-\left.
   \frac{g_{t\V}}{g_{\V\V}}\right|_{r_{H}}=\frac{a\Xi}{r_{H}^2+a^2} 
\ , \label{f6}\\ 
\frac{2\pi}{\beta_{H}}&=&
\left. \frac{\partial_{r}\Delta}{2(r^2+a^2)}\right|_{r_{H}}=
\frac{r_{H}}{2(r_{H}^2+a^2)}
\left( 1+3r_{H}^2\ell^{-2}+a^2\ell^{-2}-\frac{a^2}{r_{H}^2} \right)\ .  
\label{f7}
\end{eqnarray}
The angular momentum $\Omega_{H}$ is the same form as that given by 
Hawking, Hunter and Taylor-Robinson. 
Other definition of angular momentum given by 
Gibbons, Perry and Pope will be discussed in the next section.   
The entropy for the quantum scalar field is given by 
\begin{eqnarray} 
S_{D=4}=\frac{1}{360\pi}\frac{A_{H}}{\epsilon_{\rm inv}^{\ 2}}
\ , \label{f8}   
\end{eqnarray}
with the area on the horizon $A_{H}=4\pi(r_{H}^2+a^2)/\Xi$. 
The resultant area law (\ref{f8}) recovers that of the previous work 
for Kerr black hole case \cite{kenmoku:06}. 

\subsection{Kerr-AdS black holes in five dimensions}

In five-dimensional Einstein's field equation, 
the double-parameter rotating black hole solution holds, 
which is an example of multi-parameter black hole spacetime. 
The double-parameter solution is more complicated than single-parameter 
solution with the existence of the cross-metric term   
between two azimuthal angles. 

The metric of the Kerr-AdS black hole in five dimensions 
is given by Hawking, Hunter and Taylor-Robinson \cite{h:} as 
\begin{eqnarray}
ds_{D=5}^2=&-&\frac{\Delta}{\rho^2}
\left( dt-\frac{a\sin^2{\theta}}{\Xi_{a}}d\V_{a}
-\frac{b\cos^2{\theta}}{\Xi_{b}}d\V_{b} \right)^2 \NN\\
&+&\frac{\Delta_{\theta}\sin^2{\theta}}{\rho^2}
\left(a dt-\frac{(r^2+a^2)}{\Xi_{a}}d\V_{a} \right)^2 \NN\\
&+&\frac{\Delta_{\theta}\cos^2{\theta}}{\rho^2}
\left(b dt-\frac{(r^2+b^2)}{\Xi_{b}}d\V_{b} \right)^2 \NN\\
&+&\frac{\rho^2}{\Delta}dr^2+\frac{\rho^2}{\Delta_{\theta}}d\theta^2
+\frac{(1+r^2\ell^2)}{r^2\rho^2}
\left( ab\ dt \right. \NN\\
&-&\left. \frac{b(r^2+a^2)\sin^2{\theta}}{\Xi_{a}}d\V_{a} 
-\frac{a(r^2+b^2)\cos^2{\theta}}{\Xi_{b}}d\V_{b} \right)\ ,
\label{g1}
\end{eqnarray}
where  
\begin{eqnarray}
\Delta&=&\frac{1}{r^2}(r^2+a^2)(r^2+b^2)(1+r^2\ell^2)-2M \ ,\NN\\
\Delta_{\theta}&=&(1-a^2\ell^2\cos^2{\theta}-b^2\ell^2\sin^2{\theta}) 
\ , \NN\\
\rho^2&=&(r^2+a^2\cos^2{\theta}+b^2\sin^2{\theta})\ , \NN\\
\Xi_{a}&=&(1-a^2\ell^2) \ , \ \ \Xi_{b}=(1-b^2\ell^2)\ . 
\label{g2}
\end{eqnarray}
The notations $M,a,b$ and $\Lambda=-6\ell^{-2}$ denote 
the mass of black hole, two angular velocities and negative cosmological term.  
In order to apply the generalized area law (\ref{e40}), 
we rewrite the metric in the general form: 
\begin{eqnarray}
ds_{D=5}^2
&=&g_{tt}dt^2+2(g_{t\V_{a}}dtd\V_{a}+g_{t\V_{a}}dtd\V_{a})\NN\\  
 &&+g_{\V_{a}\V_{a}}d\V_{a}^2+g_{\V_{b}\V_{b}}d\V_{b}^2
+2g_{\V_{a}\V_{b}}d\V_{a}d\V_{b} \NN\\
&&
+g_{\theta\theta}d\theta^2+g_{rr}dr^2
\ ,  \label{g3}
\end{eqnarray}
where each components are 
\begin{eqnarray}
g_{tt}&=&\frac{1}{\rho^2}\left( -\Delta
+\Delta_{\theta}(a^2\sin^2{\theta}+b^2\cos^2{\theta})
+\Delta_{\ell} a^2b^2\right)\ , \NN\\
g_{t\V_{a}}&=&\frac{a\sin^2{\theta}}{\rho^2\Xi_{a}}
\left( \Delta -A(\Delta_{\theta}+\Delta_{\ell}b^2)\right)\ , \NN\\ 
g_{t\V_{b}}&=&\frac{b\cos^2{\theta}}{\rho^2\Xi_{b}}
\left( \Delta -B(\Delta_{\theta}+\Delta_{\ell}a^2)\right)\ , \NN\\ 
g_{\V_{a}\V_{a}}&=&\frac{\sin^2{\theta}}{\rho^2\Xi_{a}}
\left( -\Delta a^2\sin^2{\theta}+A^2(\Delta_{\theta}
+\Delta_{\ell}b^2\sin^2{\theta})\right)\ , \NN\\ 
g_{\V_{b}\V_{b}}&=&\frac{\cos^2{\theta}}{\rho^2\Xi_{b}}
\left( -\Delta b^2\cos^2{\theta}+B^2(\Delta_{\theta}
+\Delta_{\ell}a^2\cos^2{\theta})\right)\ , \NN\\ 
g_{\V_{a}\V_{b}}
&=&\frac{ab\sin^2{\theta}\cos^2{\theta}}{\rho^2\Xi_{a}\Xi_{b}}
\left( -\Delta+AB\Delta_{\ell}\right)\ , \label{g4} 
\end{eqnarray}
where some additional notations are defined as 
\begin{eqnarray}
A=r^2+a^2 \ , \ \ B=r^2+b^2 \ , \ \ \Delta_{\ell}=r^{-2}+\ell^{-2}
\ .  \label{g5}
\end{eqnarray}
We need the contravariant components of the metric 
with respect to time and azimuthal angle 
to obtain the angular velocity and temperature:  
\begin{eqnarray}
g^{tt}&=&\frac{1}{r^2\rho^2\Delta_{\theta}}
\left(Ab^2\cos^2{\theta}+Ba^2\sin^2{\theta}
-\ell^{-2}a^2b^2(A\cos^2{\theta}+B\sin^2{\theta})\right)\NN\\
&&-\frac{1}{r^4\rho^2\Delta}A^2B^2 \label{g6}\ , \\
g^{t\V_{a}}&=&\frac{a\Xi_{a}}{r^2\rho^2}
\left( \frac{1}{\Delta_{\theta}}
(B-\ell^{-2}b^2(A\cos^2{\theta}+B\sin^2{\theta}) 
-\frac{1}{r^2\Delta}AB^2\right)\label{g7}\ , \\
g^{t\V_{b}}&=&\frac{b\Xi_{b}}{r^2\rho^2}
\left( \frac{1}{\Delta_{\theta}}
(A-\ell^{-2}a^2(A\cos^2{\theta}+B\sin^2{\theta}) 
-\frac{1}{r^2\Delta}A^2B\right) \label{g8}\ .  
\end{eqnarray}
The detailed derivation is given in appendix D. 

The horizon $r_{H}$ is defined as a larger zero of the 
inverse metric components $1/g^{tt}$ and $1/g_{rr}$, which is 
given by the outer real root of $\Delta$: 
\begin{eqnarray}
\Delta(r=r_{H})=0 \Longleftrightarrow 
(r_{H}^2+a^2)(r_{H}^2+b^2)(1+r_{H}^2\ell^{-2})=2Mr_{H}^2, \label{g9}
\end{eqnarray}
The two angular velocities on the horizon Eq.(\ref{e11}) are obtained:   
\begin{eqnarray}
\Omega_{H_{a}}=\frac{g^{t\V_{a}}}{g^{tt}}=\frac{a\Xi_{a}}{r_{H}^2+a^2}
\ \ , \ \ 
\Omega_{H_{b}}=\frac{g^{t\V_{b}}}{g^{tt}}=\frac{b\Xi_{b}}{r_{H}^2+b^2} 
\ ,  \label{g10}
\end{eqnarray}
which coincide with those by Hawking, Hunter and Taylor-Robinson 
\cite{h:}. 
It is worthwhile to note that 
the angular velocities 
$\Omega_{{\rm H}_{a}}={g^{t\V_{a}}}/{g^{tt}}$ and/or 
$\Omega_{{\rm H}_{b}}={g^{t\V_{b}}}/{g^{tt}}$ 
do not equal to $-g_{t\V_{a}}/g_{\V{a}\V{a}}$ 
and/or $-g_{t\V_{b}}/g_{\V{b}\V{b}}$ 
generally in the multi-parameter case  
but they equal in single-parameter case 
due to the existence of $g_{\V_{a}\V_{b}}$ components. 
The inverse Hawking 
temperature on the horizon is given in Eq.(\ref{e35}) as  
\begin{eqnarray}
\frac{2\pi}{\beta_{H}}
=\left. \frac{r^2\partial_{r}{\Delta}}{2AB}\right|_{r_{H}}
=r_{H}(1+r_{H}^2\ell^{-2})\left(\frac{1}{r_{H}^2+a^2}+\frac{1}{r_{H}^2+b^2}
\right) -\frac{1}{r_{H}}\ . \label{g7}
\end{eqnarray}

The entropy of the five-dimensional Kerr-AdS black hole is obtained from the 
general expression in Eq.(\ref{e37}) as 
\begin{eqnarray}
S_{D=5}=\frac{5\zeta(5)}{2^{\,6}\pi^{\,6}} 
\frac{A_{H}}{\epsilon_{\rm inv}^{\ 3}} 
,  \label{g9}
\end{eqnarray}
where the area of the five-dimensional Kerr-AdS black hole is given by 
\begin{eqnarray}
A_{H}=\frac{2\pi^2 (r_{H}^2+a^2)(r_{H}^2+b^2)}{r_{H}\Xi_{a}\Xi_{b}} 
\ .  \label{g10} 
\end{eqnarray}

We have obtained quantum scalar field contribution to 
the entropy in the four- and five-dimensional spacetime, 
which will help to understand the cases in more higher-dimensional 
rotating black hole spacetime.  
For more higher dimensional black hole solutions, 
the additional number of rotating parameter 
can increase as the increase of each even dimension.   

\setcounter{equation}{0}
\section{Relation Between Thermodynamics for Scalar Field 
and Thermodynamics for Black Holes}

In this section 
we study the thermodynamics 
of scalar field which obtained in section 2 
according to the standard statistical method 
under general form of metrics 
and compare them with the corresponding  
black hole thermodynamics.  
 
First we 
can show that the statistical (Gibbs-Duhem) relation for 
the quantum scalar field holds as
\begin{eqnarray}
TS=U+p\bar{V}
\label{h1}\ , 
\end{eqnarray}
where the volume of the optical space $\bar{V}$ is given in
Eq.(\ref{e26}),  
the entropy $S$ in Eq.(\ref{e27}), the internal energy $U$ in Eq.(\ref{e28}) 
and the pressure $p$ is given by 
\begin{eqnarray}
p:=-\frac{\partial F}{\partial {\bar V}}
=\frac{\zeta{(D)}\Gamma(D/2)}{\pi^{D/2}\beta_{H}^{\ D}} 
\label{h2} \ .
\end{eqnarray}
The first law of thermodynamics also holds as
\begin{eqnarray}
TdS=dU+pd\bar{V}
\label{h3}\ .
\end{eqnarray}
The statistical relation (\ref{h1}) and the 
first law of thermodynamics (\ref{h2}) are related 
through scaling property. 
These thermodynamic relations hold because 
we have derived the thermal quantities 
according to the standard statistical mechanics method 
under general form of the black hole metrics. 

In order to compare the thermodynamics for the scalar field to 
those for black holes, 
we define the effective gravitational constant of the scalar field 
\begin{eqnarray}
G'=\frac{\pi^{3D/2-1}2^D\epsilon^{D-2}}{4\zeta(D)D\Gamma(D/2-1)}
\ ,   \label{h4}
\end{eqnarray}
which normalizes the entropy as $S=A_{H}/4G'$. 

We can express the thermal quantities 
in the language of black hole variables from that of 
scalar field variables if the explicit black hole solutions 
is used for the general form of the background metric.  
In four-dimensional Karr-AdS black hole case, which 
was studied in subsection 3.1, 
the statistical relation in the language of black holes 
is obtained in the form:
\begin{eqnarray}
TS=\frac{E_{H}}{2}-\Omega_{H}J+X_{H}\ell^{-2}\ , \label{h6}
\end{eqnarray}
where the black hole energy $E_{H}$, angular momentum $J$ and 
the conjugate variable to cosmological term $X_{H}$ are defined:   
\begin{eqnarray}
E_{H}=\frac{M}{\Xi G'}\ , \ \ 
 J=\frac{Ma}{\Xi^2 G'}
\ , \ \ 
X_{H}=\frac{r_{H}(r_{H}^2+a^2)}{2\Xi G'}
\ . \label{h7}
\end{eqnarray}
Temperature $T$ and entropy $S$ in Eq.(\ref{h6}) are the same as in
Eq.(\ref{h1}).  
The first law of the thermodynamics 
can also be expressed in language of black holes as 
\begin{eqnarray}
TdS=dE_{G}-\Omega_{G}dJ-X_{G}d\ell^{-2}\ , \label{h8}
\end{eqnarray}
where the black hole energy $E_{G}$, angular velocity $\Omega_{G}$ and  
conjugate variable to cosmological term $X_{G}$ are defined:  
\begin{eqnarray}  
E_{G}&=&\frac{M}{\Xi^2 G'} \ , \ \ 
\Omega_{G}=\frac{a(1+r_{H}^2\ell^{-2})}{r_{H}^2+a^2} \ , \nonumber\\
X_{G}&=&\frac{r_{H}(r_{H}^2+a^2)}{2\Xi^2 G'}
\left(1+\frac{a^2}{2r_{H}^2}(1-r_{H}^2\ell^{-2})\right)\ .  \label{h9}
\end{eqnarray}
The coefficients in front of each thermal quantities are from 
the scaling property; $1/2$ for energy, $1$ for thermal 
and angular terms and $-1$ for conjugate to cosmological
term. The differences between $E_{H}\ , \Omega_{H}\ , X_{H}$ 
in Eq.(\ref{h7}) and $E_{G}\ , \Omega_{G}\ , X_{G} $ in Eq.(\ref{h9}) 
are due to the non-linearity in scaling transformation in each quantities. 

We should note the difference between the angular 
velocity defined by Hawking,  
Hunter and Taylor-Robbinson $\Omega_{H}$ in Eq.(3.6) \cite{h:} and 
that by Gibbons, Perry and Pope $\Omega_{G}$ in Eq.(\ref{h9}) 
\cite{g:}. 
The angular velocity $\Omega_{H}$ is considered to be 
measured relative to a rotating frame at infinity, while  
the angular velocity $\Omega_{G}$ is measured   
relative to a non-rotating frame at infinity:  
\begin{eqnarray}
\Omega_{G }:=\Omega_{H }-\Omega_{\infty}
=\frac{a(1+r_{H}^2\ell^{-2})}{r_{H}^2+a^2} \ ,   
\ ,  \label{h10}
\end{eqnarray} 
with $\Omega_{\infty }=-a\ell^{-2}$ . 
Gibbons, Perry and Pope \cite{g:} stressed the importance to hold 
the first law of thermodynamics, in which $\Omega_{G}$ as well as 
$E_{G}$ are used in Eq.(\ref{f8}).
 
The naive derivation of the conjugate variable to cosmological term 
is the cosmological term in the action integration: 
\begin{eqnarray}
-\frac{1}{16\pi G}\int d^4x\sqrt{-g}2\Lambda
=\beta_{H}\frac{r_{H}(r_{H}^2+a^2)}{2\Xi G}\ell^{-2}
\ , \label{h11}
\end{eqnarray}
where the gravitational constant is recovered in Eq.(\ref{h11}), 
which is replaced by $G'$ in $X_{H}$ in Eq.(\ref{h7}).
For the case of the constant cosmological term;  
$d\ell^{-2}=0$, the first law (\ref{h8}) reduces to that 
given by Gibbons, Perry and Pope with the replacement: 
$G\rightarrow G'$. We include the cosmological term 
in order to keep the scaling transformation.  

By this correspondence, 
the thermodynamics for the quantum scalar fields around black holes 
is consistent with that for the black hole itself.

\section{Summary and Discussions}

We have studied the statistical mechanics for quantum scalar fields 
under multi-parameter rotating black hole spacetime 
in arbitrary $D$ dimensions.  
We have obtained the generalized area law for scalar field entropy 
around the black holes under the general form of the metric.  
The conditions imposed on the metric are the following. 

\begin{itemize}
\item[$\Diamond$ 1]
Off-diagonal components between time and azimuthal angles 
$g_{t\V_{a}}$
and among azimuthal angles $g_{\V_{a}\V_{b}}$ are assumed to exist.  
This condition indicates multi-parameter rotations with velocities:
$\Omega_{H_{a}}=g^{t\V_{a}}/g^{tt}$.

\item[$\Diamond$ 2]
Any metric components are assumed not to be functions of $t$ and 
$\V_{a}$. 
This condition ensures the energy and angular momentum conservations.

\item[$\Diamond$ 3] 
The inverse of metric components; $1/g^{tt}$ and $1/g_{rr}$ 
are assumed to have simple zeros at horizon position $r_{H}$. 

\item[$\Diamond$ 4] 
The product of the inverse metric components; 
$\partial_{r}(1/g^{tt})\times \partial _{r}(1/g_{rr})$ 
is assumed not to depend on angle $\theta$ on the horizon $r_{H}$. 
This condition leads the well-defined temperature on $r_{H}$. 
\end{itemize}

The contribution to the scalar field entropy 
is extremely dominant near the horizon in the optical volume. 
We introduce regularization parameter $\epsilon_{\rm inv}$ in the Rindler 
spacetime in order to control the divergence on the horizon.   
The large contribution near the horizon leads the area law 
of the entropy for the scalar fields naturally. 

The thermodynamics for the scalar fields can be expressed 
in the language of the black hole variables, where 
the gravitational constant is replaced by effective one of the 
scalar field; $G \rightarrow G'$.  

In deriving the area law for the quantum scalar field 
under the multi-parameter rotating black hole spacetime, 
we have postulated the quasi-equilibrium state for quantum scalar
fields.   
In connection with this, 
we should comment on the super-radiant instability for 
Kerr black hole spacetime in (3+1) dimensions 
\cite{whi:,muk:,car:}. 
Super-radiant effects occur also in our analysis through  
the energy restriction, because scalar field energy can be negative 
if the scalar field angular momentum is in the inverse direction 
with respect to the direction of the black hole angular momentum.    
The relation between our quasi-equilibrium treatment and the super-radiant 
instability phenomena will be studied in our future work.  

The quantum effects of gravity near the horizon 
is very important in deriving the black hole entropy 
and they will be our future problem too.   
\\
\vspace{\baselineskip}

\noindent
{\Large \bf Acknouledgements}\\

The authors wish to thank Dr. K. Shigemoto for his 
careful reading of the manuscript and useful comments. 
\\
\vspace{\baselineskip}

\noindent
{\Large \bf Appendix}
\appendix
\setcounter{equation}{0}
\section{Derivation of Null Potency of $\eta$ in Eq.(\ref{e10})}

In this appendix, we calculate the square of newly introduced 
Killing vector 
$\eta(=\xi_{t}+\sum_{a=1}^{p}\Omega_{H_{a}}\xi_{\V_{a}})$  
and show its null potency. 
First we note the orthogonality relation among metrics: 
\begin{eqnarray}
g_{tt}g^{tt}+\sum_{a=1}^{p}g_{t\V_{a}}g^{t\V_{a}}=1\ , \ 
g_{t\V_{a}}t^{tt}+\sum_{b=1}^{p}g_{\V_{a}\V_{b}}g^{t\V_{b}}=0\ .
\end{eqnarray}
Using orthogonality relations, we obtain the metric-angular velocity
relations: 
\begin{eqnarray}
&&g_{tt}+2\sum_{a=1}^{p}g_{t \V_{a}}\Omega_{a}+
\sum_{a,b=1}^{p}g_{\V_{a}\V_{b}}\Omega_{a}\Omega_{b}=\frac{1}{g^{tt}}\ ,\\
&&-2\sum_{a=1}^{p}g_{t\V_{a}}\Omega_{a}
-\sum_{a,b=1}^{p}g_{\V_{a}\V_{b}}\Omega_{a}\Omega_{b}
=\sum_{a,b=1}^{p}g_{\V_{a}\V_{b}}\Omega_{a}\Omega_{b}\ ,\\
&&\sum_{a=1}^{p}g_{t\V_{a}}\Omega_{{\rm H}{a}}
=-\sum_{a,b=1}^{p}g_{\V_{a}\V_{b}}\Omega_{{\rm H}{a}}\Omega_{b}
\end{eqnarray}
Then we evaluate square of $\eta$, in the following, as
\begin{eqnarray}
\eta^2
&=& 
g_{tt}+2\sum_{a=1}^{p}g_{t\V_{a}}\Omega_{{\rm H}{a}}
+\sum_{a,b=1}^{p}g_{\V_{a}\V_{b}}\Omega_{{\rm H}{a}}\Omega_{{\rm H}{b}}
\NN\\
&=& 
g_{tt}+2\sum_{a=1}^{p}g_{t\V_{a}}\Omega_{{\rm H}{a}}
+\sum_{a,b=1}^{p}g_{\V_{a}\V_{b}}\Omega_{{\rm H}{a}}\Omega_{{\rm H}{b}}
\NN\\
&+&2\sum_{a=1}^{p}g_{t\V_{a}}\Omega_{a}
+\sum_{a,b=1}^{p}g_{\V_{a}\V_{b}}\Omega_{a}\Omega_{b}
-2\sum_{a=1}^{p}g_{t\V_{a}}\Omega_{a}
-\sum_{a,b=1}^{p}g_{\V_{a}\V_{b}}\Omega_{a}\Omega_{b}
\NN\\
&=&
\frac{1}{g^{tt}}
+\sum_{a,b=1}^{p}g_{\V_{a}\V_{b}}\Omega_{a}\Omega_{b}
-2\sum_{a,b=1}^{p}g_{\V_{a}\V_{b}}\Omega_{{\rm H}{a}}\Omega_{b}
+2\sum_{a,b=1}^{p}g_{\V_{a}\V_{b}}\Omega_{{\rm H}{a}}\Omega_{{\rm H}b}
\NN\\ 
&=&
\frac{1}{g^{tt}}
+\sum_{a,b=1}^{p}g_{\V_{a}\V_{b}}(\Omega_{{\rm H}{a}}-\Omega_{a})
(\Omega_{{\rm H}{b}}-\Omega_{b})\ ,
\end{eqnarray}
where we have used the metric-angular velocity relation 
of Eqs.(A.2)-(A.4)
in the second equality to show the third equality. 
The expression of ${\eta}^2$ in Eq.(\ref{e10}) is 
the last equality in Eq.(A.5), 
which shows the null potency at horizon position 
$r=r_{H}$, where $\Omega_{{\rm H}a}=\Omega_{a}$ and $1/g^{tt}=0$. 

\setcounter{equation}{0}
\section{Derivation of Positivity of 
$\mathcal{H}'$ in Eq.(\ref{e4})}
 
In order to show the positivity of newly introduced energy density 
$\mathcal{H}'$ of Eq.(\ref{e13}), 
we insert the explicit expression of 
scalar fields in the Lagrangian density (\ref{e4}) 
and rewrite as  
\begin{eqnarray}
{\mathcal{H}}'
&=&
\Pi\partial_{t}\Phi -{\mathcal{L}}_{\rm scalar}
+\sum_{a=1}^{p}
\Omega_{{\rm H}a}\Pi\partial_{\V_{a}}\Phi\NN\\
&=&
\Pi\partial_{t}\Phi 
+\frac{1}{2}(g^{tt}(\partial_{t}\Phi)^2 
+2\sum_{a=1}^{p}g^{t\V_{a}}\partial_{t}\Phi\partial_{\V_{b}}\Phi 
+\sum_{a,b=1}^{p}g^{\V_{a}\V_{b}}
\partial_{\V_{a}}\Phi\partial_{\V_{b}}\Phi \NN\\
&&+\left. (r,\theta \ {\mbox{and mass terms}})\right)
+\sum_{a=1}^{p}
\Omega_{{\rm H}a}\Pi\partial_{\V_{a}}\Phi\ \NN\\
&=&
\frac{1}{2}(
-g^{tt}(\partial_{t}\Phi)^2 
+\sum_{a,b=1}^{p}
g^{\V_{a}\V_{b}}\partial_{\V_{a}}\Phi\partial_{\V_{b}}\Phi 
+(r,\theta \ {\mbox{and mass terms}})) \NN\\
&&+\sum_{a=1}^{p}\Omega_{{\rm H}a}\Pi\partial_{\V_{a}}\Phi \ . 
\end{eqnarray}
After eliminate the time derivative terms by means of the momentum 
(\ref{e5}),  we obtain 
\begin{eqnarray}
{\mathcal{H}}'
&=&
\frac{1}{2}(
-\frac{1}{g^{tt}}(\Pi+\sum_{a=1}^{p}g^{t\V_{a}}\partial_{{\V_{a}}}\Phi)^2  
+\sum_{a,b=1}^{p}
g^{\V_{a}\V_{b}}\partial_{\V_{a}}\Phi\partial_{\V_{b}}\Phi \NN\\
&&+(r,\theta \ {\mbox{and mass terms}})) 
+\sum_{a=1}^{p}\Omega_{{\rm H}a}\Pi\partial_{\V_{a}}\Phi \NN\\
&=&
\frac{1}{2}(
-\frac{1}{g^{tt}}{\Pi}^2  
+\sum_{a,b=1}^{p}
{\tilde g}^{\V_{a}\V_{b}}\partial_{\V_{a}}\Phi\partial_{\V_{b}}\Phi 
+(r,\theta \ {\mbox{and mass terms}})) \NN\\
&&+\sum_{a=1}^{p}(\Omega_{{\rm
 H}a}-\Omega_{{a}})\Pi\partial_{\V_{a}}\Phi 
\ , 
\end{eqnarray}
where ${\tilde g}^{\V_{a}\V_{b}}$ is the inverse metric 
of ${g}_{\V_{a}\V_{b}}$ defined as 
\begin{eqnarray}
{g}_{\V_{a}\V_{b}}{\tilde g}^{\V_{b}\V_{c}}=\delta_{a}^{c}\ \ , \
 \ 
{\tilde g}^{\V_{a}\V_{b}}:=
{g}^{\V_{a}\V_{b}}-\frac{g^{t\V_{a}}g^{t\V_{b}}}{g^{tt}}\ .
\end{eqnarray}
More conveniently we introduce the inverse metric $h^{ij}$ of 
special part metrics $h_{ij}$ defined by 
\begin{eqnarray}
h^{ij}:=({\tilde{g}}^{\V_{a}\V_{b}}, 
g^{\theta_{m}\theta_{m}}=\frac{1}{g_{\theta_{m}\theta_{m}}}, 
g^{rr}=\frac{1}{g_{rr}})\ ,
\end{eqnarray}
which is the expression of the inverse metric (\ref{e14}).  
The desired form for $\mathcal{H}'$ of Eq.(\ref{e13}) is obtained:   
\begin{eqnarray}
{\mathcal{H}}'
=
\frac{1}{2}(
-\frac{\Pi^2}{g^{tt}}  
+\sum_{i,j}^{D-1}
{h}^{ij}\partial_{i}\Phi\partial_{j}\Phi 
+\mu^2\Phi^2) 
+\sum_{a=1}^{p}(\Omega_{{\rm H}a}-\Omega_{{a}})\Pi\partial_{\V_{a}}\Phi 
\  ,    
\end{eqnarray}
where this new Hamiltonian density is positive definite under the 
near horizon approximation $\Omega_{{\rm H}a}\simeq \Omega_{a}$.

\setcounter{equation}{0}
\section{Derivation of Line Element in Eq.(\ref{e31})}

Using the definition of angular velocity (\ref{e11}) and 
orthogonality relation (A.1), we obtain another metric-angular 
momentum identities: 
\begin{eqnarray}
g_{tt}&=&\frac{1}{g^{tt}}
+\sum_{a,b=1}^{p}g_{\V_{a}\V_{b}}\Omega_{\V_{a}}\Omega_{\V_{b}}
\ , \NN\\
g_{t\V_{a}}&=&-\sum_{b=1}^{p}g_{\V_{a}\V_{b}}\Omega_{\V_{b}} \ .
\end{eqnarray}
Then the line element in the original form (\ref{e2}) 
is rewritten using the identities as
\begin{eqnarray}
ds^2
&=&
g_{\tau\tau}{d\tau}^2+2\sum_{a=1}^{p}g_{\tau\V_{a}}d\tau d\V_{a} 
+(r,\theta \ {\rm terms})\ , \NN\\
&=& (\frac{1}{g^{\tau\tau}}+
\sum_{a,b=1}^{p}g_{\V_{a}\V_{b}}\Omega_{\V_{a}}\Omega_{\V_{b}})d\tau^2\ 
- 2\sum_{a,b=1}^{p}g_{\V_{a}\V_{b}}\Omega_{b}dtd\V_{a} \NN\\
&&+\sum_{a,b=1}^{p}g_{\V_{a}\V_{b}}d\V_{a}d\V_{b}
+(r,\theta \ {\rm terms}) \ ,\NN\\
&=&\frac{1}{g^{\tau\tau}}d\tau^2
+\sum_{a,b=1}^{p}
g_{\V_{a}\V_{b}}(d\V_{a}-\Omega_{a}d\tau)(d\V_{b}-\Omega_{b}d\tau)\NN\\
&&
+(r,\theta \ {\rm terms}) \ ,   
\end{eqnarray}
which complete the derivation of Eq.(\ref{e31}).

\setcounter{equation}{0}
\section{Contravariant Metric Components for Kerr-AdS Black
 Holes in Five Dimensions} 

We derive the contravariant metric components  
with respect to time and azimuthal angle 
to calculate the area law of the black hole entropy, 
which are defined in general form as 
\begin{eqnarray}
g^{tt}=\frac{\gamma_{t}}{\Gamma} \ \ , \ \ 
g^{t\V_{a}}=\frac{\gamma_{\V_{a}}}{\Gamma}  \ \ , \ \ 
g^{t\V_{b}}=\frac{\gamma_{\V_{b}}}{\Gamma} \ , \label{d1}
\end{eqnarray}
where 
\begin{eqnarray}
\gamma_{t}&=&g_{\V_{a}\V_{a}}g_{\V_{b}\V_{b}}-g_{\V_{a}\V_{b}}^2 \NN\\
\gamma_{\V_{a}}&=&g_{t\V_{b}}g_{\V_{a}\V_{b}}-g_{t\V_{a}}g_{\V_{b}\V_{b}}\NN\\
\gamma_{\V_{b}}&=&g_{t\V_{a}}g_{\V_{a}\V_{b}} -g_{t\V_{b}}g_{\V_{a}\V_{a}}
\label{d2} ,   
\end{eqnarray}
and the determinant of the time and azimuthal angle components is
defined as 
\begin{eqnarray}
\Gamma
&=&g_{tt}g_{\V_{a}\V_{a}}g_{\V_{b}\V_{b}}
+2g_{t\V_{a}}g_{t\V_{b}}g_{\V_{a}\V_{b}} 
-g_{t\V_{a}}^2g_{\V_{b}\V_{b}}
-g_{t\V_{b}}^2g_{\V_{a}\V_{a}}
-g_{\V_{a}\V_{b}}^2g_{tt}
 \NN\\
&=&g_{tt}\gamma_{t}+g_{t\V_{a}}\gamma_{\V_{a}}+g_{t\V_{b}}\gamma_{\V_{b}}
\label{d3}\ ,  
\end{eqnarray}
Inserting the five-dimensional Kerr-AdS black hole solution (\ref{g4}), 
the factor $\gamma$'s (\ref{g2}) are expressed:  
\begin{eqnarray}
\gamma_{t}&\times&\frac{r^2\rho^2\Xi^2_{a}\Xi^2_{b}}
{\sin^2{\theta}\cos^2{\theta}}\NN\\
&=& r^2\Delta\left(-Ab^2\cos^2{\theta}-Ba^2\sin^2{\theta}
+\ell^{-2}a^2b^2(A\cos^2{\theta}+B\sin^2{\theta})\right)\NN\\
&&+ \Delta_{\theta}A^2B^2 \ , \label{d4}\\
\gamma_{\V_{a}}&\times&\frac{r^2\rho^2\Xi_{a}\Xi^2_{b}}
{a\sin^2{\theta}\cos^2{\theta}}\NN\\
&=&
 r^2\Delta\left(-B+\ell^{-2}b^2(A\cos^2{\theta}+B\sin^2{\theta})\right)
+ \Delta_{\theta}AB^2 \label{d5}\ , \\
\gamma_{\V_{b}}&\times&\frac{r^2\rho^2\Xi_{a}^2\Xi_{b}}
{b\sin^2{\theta}\cos^2{\theta}}\NN\\
&=&
 r^2\Delta\left(-A+\ell^{-2}a^2(A\cos^2{\theta}+B\sin^2{\theta})\right)
+ \Delta_{\theta}A^2B \label{d6}\ , 
\end{eqnarray}
and $\Gamma$ of Eq.(\ref{d3}) is expressed:  
\begin{eqnarray}
\Gamma=-\frac{\sin^2{\theta}\cos^2{\theta}r^2\Delta\Delta_{\theta}}
{\Xi^2_{a}\Xi^2_{b}} \ .  \label{d7} 
\end{eqnarray}
The contravariant metric of time and azimuthal components 
are obtained using Eqs.(\ref{d1})-(\ref{d7}): 
\begin{eqnarray}
g^{tt}&=&\frac{1}{r^2\rho^2\Delta_{\theta}}
\left(Ab^2\cos^2{\theta}+Ba^2\sin^2{\theta}
-\ell^{-2}a^2b^2(A\cos^2{\theta}+B\sin^2{\theta})\right)\NN\\
&&-\frac{1}{r^4\rho^2\Delta}A^2B^2 \label{d8}\ , \\
g^{t\V_{a}}&=&\frac{a\Xi_{a}}{r^2\rho^2}
\left( \frac{1}{\Delta_{\theta}}
(B-\ell^{-2}b^2(A\cos^2{\theta}+B\sin^2{\theta}) 
-\frac{1}{r^2\Delta}AB^2\right)\label{d9}\ , \\
g^{t\V_{b}}&=&\frac{b\Xi_{b}}{r^2\rho^2}
\left( \frac{1}{\Delta_{\theta}}
(A-\ell^{-2}a^2(A\cos^2{\theta}+B\sin^2{\theta}) 
-\frac{1}{r^2\Delta}A^2B\right) \label{d10}\ ,   
\end{eqnarray}
which are the expressions (\ref{g6})-(\ref{g8}).


\end{document}